\documentclass[letterpaper, 10 pt, conference]{ieeeconf}  
\IEEEoverridecommandlockouts 
\IEEEoverridecommandlockouts                              

\makeatletter
\let\proof\@undefined
\let\endproof\@undefined
\makeatother
\usepackage{amsmath}
\usepackage{cite} 
\usepackage{array}
\usepackage{graphicx}
\usepackage{tcolorbox}
\usepackage{xcolor}
\usepackage{engord}
\usepackage{bm}
\usepackage{booktabs}
\usepackage{amssymb}
\usepackage{multirow}
\usepackage{longtable}
\usepackage{rotating}
\usepackage{float}
\usepackage{amsthm}
\usepackage{bm}
\usepackage{mathtools}
\usepackage[font=footnotesize]{caption}
\usepackage{hyperref}
\usepackage[ruled]{algorithm2e} 
\newtheoremstyle{nonboldnonitalic}  
  {}  
  {}  
  {\normalfont}  
  {}  
  {\itshape}  
  {.}  
  {.5em}  
  {}  

\theoremstyle{nonboldnonitalic}
\newtheorem{theorem}{Theorem}

\newtheorem{definition}[theorem]{Definition}
\newtheorem{proposition}[theorem]{Proposition}

\theoremstyle{remark}

\newcommand{\revised}[1]{{\color{black}#1}}

\let\emptyset\varnothing
\newcommand{\mb}[1]{\mathbf{#1}}
\newcommand{\mbg}[1]{\boldsymbol{#1}}
\newcommand{\minimize}[1]{\underset{{#1}}{\text{minimize}}}
\newcommand{\maximize}[1]{\underset{{#1}}{\text{maximize}}}

\newcommand{\st}{\text{subject to}}
\newcommand{\theoremsymbol}{\hfill \ensuremath{\triangleleft}}

\title{\LARGE \bf
Synthesizing Grid Data with Cyber Resilience and Privacy Guarantees
}

\author{Shengyang Wu and Vladimir Dvorkin
\thanks{Shengyang Wu ({\tt syseanwu@umich.edu}) and Vladimir Dvorkin ({\tt dvorkin@umich.edu}) are with the Department of Electrical Engineering and Computer Science, University of Michigan, MI 48109, USA.}
}

\begin{document}
\begingroup
\allowdisplaybreaks

\maketitle
\thispagestyle{empty}
\pagestyle{empty}

\begin{abstract}
Differential privacy (DP) provides a principled approach to synthesizing data (e.g., loads) from real-world power systems while limiting the exposure of sensitive information. However, adversaries may exploit synthetic data to calibrate cyberattacks on the source grids. To control these risks, we propose new DP algorithms for synthesizing data that provide the source grids with both cyber resilience and privacy guarantees. The algorithms incorporate both normal operation and attack optimization models to balance the fidelity of synthesized data and cyber resilience. The resulting post-processing optimization is reformulated as a robust optimization problem, which is compatible with the exponential mechanism of DP to moderate its computational burden.
\end{abstract}

\section{Introduction}

Optimal power flow (OPF) analysis in power systems requires realistic grid models with accurate network, generation, and load parameters—data that is difficult to source from real-world grids due to privacy and \revised{(cyber-)security} concerns. While the lack of such models has inspired the development of artificial grids \cite{8333771,Taylor2024}, a more principled approach leverages the theory of differential privacy (DP) \cite{Dwork2014} to release grid models directly from real-world systems.

The DP theory asserts that it is impossible—up to prescribed privacy parameters—to infer the original parameters from their DP release. Such strong privacy guarantees originate from Laplacian perturbations \cite{dwork2006calibrating} of real grid parameters, followed by post-processing optimization of the perturbed parameters to restore their modeling fidelity to the source grid, e.g., in terms of similarity of the OPF outcomes \cite{Fioretto2020,Mak2020,Dvorkin2023}. The DP theory also lies at the core of modern privacy-preserving OPF solvers \cite{Dvorkin2020,Dvorkin2021,Ryu2022}, the release of aggregated grid statistics \cite{Zhou2019}, and related grid information \cite{Ravi2022}.

However, the privacy guarantees alone may not suffice to release  grid parameters, as cybersecurity risks associated with such releases remain largely unexplored. Possible cyber attacks include \textit{false data injection}, which subtly alters state estimation results \cite{Liu2011}, \revised{\textit{line
outage masking}, which disconnects a transmission line and misguides a control center to seek outage elsewhere \cite{Chung2019}}, and \textit{load redistribution}, which manipulates demand measurements to increase OPF cost and constraint violation \cite{Liu2014}. \revised{The latter is of main interest to this work.} Executing such attacks requires some grid knowledge \cite{Liang2016}, which is traditionally difficult to obtain. However, the availability of synthetic grid data may unintentionally inform adversaries and help them calibrate the attack.

\textit{Contribution:} Recognizing the risks that synthetic grid parameters may inform cyber adversaries, we develop new DP algorithms that simultaneously guarantee cyber resilience and privacy for the source power grids. Our algorithms build on \cite{Fioretto2020, Mak2020, Dvorkin2023} and leverage the Laplace mechanism and post-processing optimization to tune synthetic data while anticipating cyber risks through embedded attack optimization. 

The contributions of this paper are summarized as follows:
\begin{enumerate}
    \item We formulate a Cyber Resilient Obfuscation (CRO) algorithm, an optimization-based algorithm to release electric load data with a guarantee to preserve the privacy of the original data and ensure the resilience of the source grid to load redistribution attacks. The algorithm post-processes synthetic loads by balancing their fidelity with the potential damage to the grid.
    
    \item The underlying post-processing optimization is an intractable trilevel problem, which is reduced to a tractable yet more conservative bilevel problem. We achieve this by exploring the connections between robust and bilevel optimization, in the spirit of \cite{goerigk2024connections}.
    
    \item To further improve computational tractability of the algorithm, we provide the extension of CRO, termed CRO-Exp, which uses the exponential mechanism of DP to identify only the most important constraints for post-processing optimization of synthetic loads.
\end{enumerate}

We next provide preliminaries on OPF and DP theory. Sec. \ref{sec:DP_cyber_risks} explains the risks of cyberattacks, and Sec. \ref{sec:main} introduces the algorithms to mitigate them. Sec. \ref{sec:experiments} presents simulations, and Sec. \ref{sec:conclusions} concludes. Proofs are relegated to the appendix.

\textit{Notation:} lower- (upper-) case boldface letters denote column vectors (matrices). Scalar $a_{i}$ is the $i^{\text{th}}$ element of vector $\mb{a}$. Vectors $\mb{0}$ and $\mb{1}$ are the all-zero and all-one vectors; $\top$ stands for transposition, and $\mb{x}^\star$ is the optimal value of $\mb{x}$.

\section{Preliminaries}
\subsection{\revised{DC Optimal Power Flow (OPF)} Problem}
\revised{For a given load vector $\mb{d}$, the DC OPF problem seeks the least-cost generation dispatch in high-voltage grids that satisfies the loads and grid limits. Consider OPF as a parametric linear program:}
        \begin{subequations}\label{eq:DCOPF}
        \begin{align}
            C_{\text{opf}}(\mb{d})=\minimize{\mb{p},\mb{v}\geqslant\mb{0}}\quad& \mb{q}^{\top}\mb{p}+\mbg{\psi}^{\top}\mb{v}\\
            \st\quad
            & \revised{\underline{\mb{p}}} \leqslant \mb{p} \leqslant \overline{\mb{p}} \label{eq:DCOPF pg}\\
            &\bm{1}^{\top}(\mb{p}-\mb{d})=0 \label{eq:DCOPF pb}\\
            &|\mb{F}(\mb{p}-\mb{d})|\leqslant \overline{\mb{f}}+\mb{v} \label{eq:DCOPF lcu}
        \end{align}
        \end{subequations}
        \revised{where decision variables include generator dispatch $\mb{p}$, bounded by dispatch range $[\underline{\mb{p}}, \overline{\mb{p}}]$, and power flow constraint violations $\mb{v}$, penalized by $\mbg{\psi}$. The matrix $\mb{F}$ of power transfer distribution factors is used to map net power injections $(\mb{p}-\mb{d})$ to power flows as $\mb{F}(\mb{p}-\mb{d})$.} Constraint \eqref{eq:DCOPF pb} defines system-wide power balance between dispatched generation and loads. The power flows in transmission lines are capped by line capacity $\overline{\mb{f}}$ using constraint \eqref{eq:DCOPF lcu}. In highly loaded condition, these constraints can be temporally violated by $\mb{v}\geqslant\mb{0}$. As transmission constraint violations are not desired, they are penalized with a large parameter $|\mbg{\psi}|\gg|\revised{\mb{q}}|$.

We write the linear OPF problem \eqref{eq:DCOPF} in a compact form
\begin{subequations}\label{eq:DCOPFc}
\begin{align}
    C_{\text{opf}}(\mb{d})=\minimize{\mb{x}}\quad& \mb{c}^{\top}\mb{x}\\
    \st\quad& \mb{a}_{k}^{\top}\mb{x} + \mb{b}_{k}^{\top}\mb{d} + e_{k} \leqslant \mb{0}, \nonumber\\
    &\quad\quad\quad\!\; \forall k=1,\dots,K, \label{eq:DCOPFc_con}
\end{align}
\end{subequations}
where $\mb{x} = [\mb{p}^{\top}\;\mb{v}^{\top}]^{\top}$, $\mb{c} = [\mb{q}^{\top}\;\mbg{\psi}^{\top}]^{\top}$. The $K$ inequalities in \eqref{eq:DCOPFc_con} encode the dispatch constraints \eqref{eq:DCOPF pg}, power flow constraints \eqref{eq:DCOPF pb} and \eqref{eq:DCOPF lcu} using properly specified parameters $\mb{a}_{1},\mb{b}_{1},e_{1},\dots,\mb{a}_{K},\mb{b}_{K},e_{K}$. 


\subsection{Differential Privacy for Synthetic OPF Datasets}\label{subsec:DP_Synth_OPF}

Optimization parameters in problem \eqref{eq:DCOPFc} are either classified or owned by private system actors, and thus can not be directly disclosed to public. Our goal is thus to {\it synthesize} some realistic version of these parameters. \revised{In this work,} we focus on the obfuscation of demand vector $\mb{d}$.  \revised{This is without much loss of generality, because other parameters, such as transmission data in $\mb{a}_k$, $\mb{b}_k$ and $e_{k}$, can be synthesized similarly; see the state-of-the-art obfuscation algorithms \cite{Fioretto2020,Mak2020,Dvorkin2023}}. \revised{Towards this goal,} \revised{we leverage DP to render the original vector $\mb{d}$ statistically indistinguishable from its synthetic counterpart $\tilde{\mb{d}}$, up to some prescribed parameters: $\alpha$, termed the \textit{adjacency} parameter, and $\varepsilon$, termed the \textit{privacy loss} \cite{Dwork2014}.}
\begin{definition}[Adjacency]\label{def:adjacency} Two vectors $\mb{d},\mb{d}'\in\mathcal{D}\subset\mathbb{R}^{n}$ are $\alpha-$adjacent, \revised{for some $\alpha>0$}, if \revised{$\exists i\in\{1,\dots,n$\}}, such that $d_{j}=d_{j}',\forall j\in\{1,\dots,n\}\backslash i$, and $|d_{i}-d_{i}'|\leqslant\alpha$. That is, they are different in one item by at most $\alpha$. \theoremsymbol
\end{definition}

To synthesize a DP version $\tilde{\mb{d}}$ of $\mb{d}$, the standard Laplace mechanism applies a random noise to the original data, i.e., $\tilde{\mb{d}} = \mb{d} + \text{Lap}(\frac{\alpha}{\varepsilon})^{n}$, where $\text{Lap}(s)^{n}$ is \revised{a} random draw from the $n-$dimensional Laplace distribution with zero mean \revised{and diagonal covariance matrix with each diagonal element equal $2s^2$} \cite{Mak2020}. \revised{The mechanism guarantees that if the attacker’s prior for any load is within the $\pm\alpha$ MW range of the true value, it will not be improved by the DP release. If the prior is outside this range, the prior knowledge will be improved (thus enhancing grid transparency), but the exact loads will not be disclosed. In other words,} the mechanism satisfies the following definition of $\varepsilon-$DP.
\begin{definition}[$\varepsilon-$DP]\label{def:epsilon} 
    \revised{The Laplace mechanism above, with domain $\mathcal{D}$ and output range $\mathcal{O}$, is called $\varepsilon-$DP if, for any outcome within $\hat{\mathcal{O}}\subseteq \mathcal{O}$} and any two $\alpha-$adjacent load vectors $\mb{d}$ and $\mb{d}'$, the ratio of probabilities is bounded as
    \revised{\begin{align}
        \frac{\text{Pr}[\mb{d}\textcolor{white}{'} + \text{Lap}(\frac{\alpha}{\varepsilon})^{n}\in \hat{\mathcal{O}}]}{\text{Pr}[\mb{d}' + \text{Lap}(\frac{\alpha}{\varepsilon})^{n}\in \hat{\mathcal{O}}]}\leqslant\text{exp}(\varepsilon).
    \end{align}} where $\varepsilon$ is a prescribed non-negative parameter.
\theoremsymbol
\end{definition}
Intuitively, a smaller privacy loss $\varepsilon$ results in more noise applied to data and higher requirement for distribution similarity, which would make it more likely to observe the same random outcome. 
However, the Laplace mechanism alone is likely to produce such load vector $\tilde{\mb{d}}$ that does not admit a feasible OPF solution, i.e.,  $C_{\text{opf}}(\tilde{\mb{d}})=\emptyset.$
The prior work introduced the following two-stage solution:  
\begin{enumerate}
    \item Laplace mechanism $\tilde{\mb{d}}^{0}=\mb{d} + \text{Lap}\left(\!\frac{2\alpha}{\varepsilon}\!\right)^{n}$, followed by
    \item Post-processing of $\tilde{\mb{d}}^{0}$ using a bilevel optimization: 
\end{enumerate}
\begin{align}\label{eq:blpp}
    \minimize{\tilde{\mb{d}}}\;\;& \lVert C_\text{opf}(\tilde{\mb{d}})-\tilde{C}_\text{opf} \rVert_1 +\gamma \lVert \tilde{\mb{d}}-\tilde{\mb{d}}^0 \rVert_1
\end{align}
\revised{where the OPF costs $C_\text{opf}(\tilde{\mb{d}})$ comes from the embedded optimization problem \eqref{eq:DCOPFc} formulated on synthetic load vector $\tilde{\mb{d}}$, and $\tilde{C}_\text{opf}=C_\text{opf}(\mb{d})+\text{Lap}(\frac{2\overline{c}}{\varepsilon})$ computes a DP estimate of OPF costs on true data with $\overline{c}$ being the cost of the most expensive generator.}
The synthetic vector $\tilde{\mb{d}}$ is optimized using feedback from the embedded OPF problem, which constraints $\tilde{\mb{d}}$ to take only those values that admit a feasible OPF solution. The main objective of \eqref{eq:blpp} is to match the OPF cost on synthetic load vector with that on the original load vector, thereby ensuring high modeling fidelity of the synthetic data. \revised{The second term in \eqref{eq:blpp} is a regularization term with some small hyper-parameter $\gamma>0$ to choose the optimal solution that is closest to the original load after DP obfuscation $\tilde{\mb{d}^0}$.} Solution to \eqref{eq:blpp} is the feasible and cost-consistent synthetic counterpart $\tilde{\mb{d}}$, which ensures $\varepsilon-$DP guarantee for the original load vector $\mb{d}$ \cite{Dvorkin2023}. 

One barrier to releasing synthetic OPF parameters is the risk posed by cyber adversaries who might exploit them to disrupt grid operations. Next, we substantiate these risks.

\section{Cyber Resilience Risks in Releasing Differentially Private OPF Datasets}\label{sec:DP_cyber_risks}

Although synthetic OPF datasets contribute to overall grid transparency and enable independent power flow analysis, they can also be misused by cyber adversaries launching attacks on the grid. One class of attacks, which is of interest to this work, is \textit{load redistribution} attacks. In terms of OPF problem \eqref{eq:DCOPFc}, the adversary optimizes an attack vector $\mbg{\delta}$ that alters loads in $\mb{d}$ to increase either the dispatch cost or the magnitude of power flow constraint violations. 

According to \cite{Liu2014}, the optimal attack vector is found by solving the following bilevel optimization (BO) problem: 
\begin{subequations}\label{eq:BiAttack}
\begin{align}
    C_{\text{att}}^\text{BO}(\mb{d})= \maximize{\bm{\delta}\in \Delta}\; C_{\text{opf}}(\mb{d}+\bm{\delta})
\end{align}
\revised{where the OPF costs $C_\text{opf}(\mb{d}+\mbg{\delta})$ comes from the embedded optimization problem \eqref{eq:DCOPFc} formulated on load vector after attack $\mb{d}+\mbg{\delta}$.} The attack vector is constrained by the set of admissible attacks 
\begin{align}\label{eq:BiAttack st1}
    \Delta&\triangleq
    \left\{ \bm{\delta} \;\bigg|
    \begin{array}{l}
       \underline{\bm{\delta}}\leqslant\bm{\delta}\leqslant\overline{\bm{\delta}} \\
       \bm{1}^{\top}\bm{\delta} = 0 
    \end{array}
    \right\} 
\end{align}
\end{subequations}
where $\overline{\bm{\delta}}$ and $\underline{\bm{\delta}}$ are limits on attack magnitude, and $\bm{1}^{\top}\bm{\delta} = 0$ ensures that the total system loading remains unchanged after the attack, thus ensuring the stealthiness of the attack. 

While the actual load vector $\mb{d}$ is not revealed to public, the adversary may leverage its DP release $\tilde{\mb{d}}$ to calibrate the attack. Our experiments in Sec. \ref{sec:experiments} reveal that the vector computed on $\tilde{\mb{d}}$ leads to a substantial increase of OPF costs across standard power systems benchmarks (see Tab. \ref{tab:BO att}).

\section{Cyber Resilience and Privacy Guarantees for Synthetic OPF Datasets}\label{sec:main}

Recognizing the risks of misusing synthetic datasets, we revisit the post-processing to enhance the cyber resilience of source grids. Instead of \eqref{eq:blpp}, we propose the following upper-level objective for the post-processing optimization: 
\begin{align}\label{eq:trilevel}
    \minimize{\tilde{\mb{d}}}\quad \lVert C_\text{opf}(\tilde{\mb{d}}) - \tilde{C}_\text{opf}  \rVert_1 &+ \beta\lVert C_\text{att}^{\text{BO}}(\tilde{\mb{d}})-\tilde{C}_\text{opf} \rVert_1 \nonumber\\[-5pt]
    &\quad\revised{+\gamma \lVert \tilde{\mb{d}}-\tilde{\mb{d}}^0 \rVert_1}
\end{align}
where the first term \revised{controls} the fidelity of the synthetic data, the second term measures the damage under attack calibrated on the synthetic data, \revised{and the third term regularizes the demand vector}. \revised{For a small penalty $\gamma$, this objective represents a trade-off between} the fidelity of synthetic grid parameters and resilience of the grid to redistribution attacks, which can be explored by varying parameter $\beta>0.$ \revised{The embedded optimization $C_\text{att}^{\text{BO}}(\tilde{\mb{d}})$ includes the real grid data except for the load vector, thus modeling the worst-case attack when only the loads are unknown to adversaries.}

The challenge is that \eqref{eq:trilevel} requires solving a \textit{trilevel} optimization problem, where the synthetic data is optimized over embedded BO attack model $C_\text{att}^{\text{BO}}(\tilde{\mb{d}})$. Inspired by \cite{goerigk2024connections}, we seek computational tractability by exploring the connection between the bilevel model of attack and robust optimization.

\subsection{Computational Tractability via Robust Optimization (RO)}

The conservative RO approximation of \eqref{eq:BiAttack} is
\begin{subequations}\label{prob:RO}
\begin{align}
    C_{\text{att}}^{\text{RO}}(\mb{d})=&\;\minimize{\mb{x}}\;\; \mb{c}^{\top}\mb{x}\\
    \st\;\;\;& \underset{\mbg{\delta_k\in\Delta}}{\max}\left[\mb{a}_{k}^{\top}\mb{x} + \mb{b}_{k}^{\top}(\mb{d}+\bm{\delta}_k) + e_{k}\right] \leqslant \mb{0},\forall k,\!\!\!
\end{align}
\end{subequations}
where each constraint $k$ is formulated for the worst-case realization of the attack vector from the set of admissible attacks. In contrast to bilevel formulation \eqref{eq:BiAttack}, the RO attack \revised{generates a worst-case attack vector for} each constraint. The following result shows that the RO attack provides an upper-bound on the BO attack. 

\begin{proposition}[Conservative attack approximation]\label{th:att} For any feasible load vector $\mb{d}$, relation $C_{\text{att}}^{\text{RO}}(\mb{d}) \geqslant C_{\text{att}}^\text{BO}(\mb{d})$ holds. \theoremsymbol  
\end{proposition} 

Although conservative, formulation \eqref{prob:RO} is computationally advantageous over \eqref{eq:BiAttack}  as it admits a linear programming reformulation via duality \revised{\cite[\S 2.2]{Bertsimas2011}} \revised{(see the link to online repository below for details)}. Let $\underline{\mbg{\mu}}$ and $\overline{\mbg{\mu}}$ be the duals of the first constraints in \eqref{eq:BiAttack st1}, and $\lambda$ be the dual of the last condition in \eqref{eq:BiAttack st1}. The exact reformulation of \eqref{prob:RO} is
\begin{subequations}\label{prob:RO_ref}
\begin{align}
    C_{\text{att}}^{\text{RO}}(\mb{d})=\;\minimize{\mb{x},\underline{\bm{\mu}},\overline{\bm{\mu}},\lambda}\;\;\;& \mb{c}^{\top}\mb{x}\\
    \st\;\;\;& \mb{a}_{k}^{\top}\mb{x} + \mb{b}_{k}^{\top}\mb{d}\nonumber\\
    &\;+\overline{\bm{\mu}}_k^{\top}\overline{\bm{\delta}}-\underline{\bm{\mu}}_k^{\top}\underline{\bm{\delta}} + e_{k} \leqslant \mb{0},\!\!\!\\
    &\mb{b}_k-\overline{\bm{\mu}}_k+\underline{\bm{\mu}}_k-\bm{1}\lambda_k=\bm{0},\\
    &\underline{\bm{\mu}}_{k},\overline{\bm{\mu}}_{k} \geqslant\mb{0},\forall k=1,\dots,K.
\end{align}
\end{subequations}
Therefore, replacing $C_{\text{att}}^{\text{BO}}$ with $C_{\text{att}}^{\text{RO}}$ in objective function \eqref{eq:trilevel} gives rise to bilevel post-processing optimization, which can be handled by mixed-integer optimization solvers \cite{Dvorkin2023,Mak2020}.

Next, we introduce a tractable post-processing algorithm for synthesizing loads with privacy and cyber resilience guarantees. Then, in Sec. \ref{subsec:exponential_mech_extension}, we modify the algorithm to tune the computational burden of the RO approximation.

\subsection{Differentially Private CRO}
\begin{algorithm}[b]
  \small
  \KwIn{$\mb{d}$, $(\alpha,\varepsilon_1,\varepsilon_2)$, $(\beta,\gamma,\Delta)$ 
  }

  {\footnotesize \textbf{1}} Initial load obfuscation: $\tilde{\mb{d}}^{0}=\mb{d} + \text{Lap}\left(\!\frac{\alpha}{\varepsilon_1}\!\right)^{n}$

  {\footnotesize \textbf{2}}  DP estimation of OPF costs: $\tilde{C}_\text{opf}=C_\text{opf}(\mb{d})+\text{Lap}\left(\frac{\alpha\overline{c}}{\varepsilon_2}\right)$

  {\footnotesize \textbf{3}}  Post-processing optimization of the synthetic load vector
  \begin{align}\label{eq:pp_cro}
       \tilde{\mb{d}} \in \underset{\tilde{\mb{d}}}{\text{argmin}}\;& \lVert C_\text{opf}(\tilde{\mb{d}}) - \tilde{C}_\text{opf}  \rVert_1 \nonumber \\[-5pt]
       &\quad + \beta\lVert C_\text{att}^{\text{RO}}(\tilde{\mb{d}})-\tilde{C}_\text{opf} \rVert_1 + \gamma \lVert \tilde{\mb{d}}-\tilde{\mb{d}}^0 \rVert_1
    \end{align}
    
    \KwOut{Synthetic load vector $\tilde{\mb{d}}$}
  \caption{Privacy-preserving CRO}\label{alg:CRO}
\end{algorithm}

The CRO algorithm for privacy-preserving and cyber-resilient synthesis of load parameters is summarized in Alg. \ref{alg:CRO}. It takes as inputs load adjacency and $\varepsilon$-DP parameters, as well as optimization trade-off, regularization and attack parameters, $\beta$,  $\gamma$ and $\Delta$, respectively. Step 1 initializes the synthetic load vector using the Laplace mechanism with a privacy loss of $\varepsilon_1$. Step 2 performs a DP estimation of the OPF costs on real loads using the Laplace mechanism with a privacy loss of $\varepsilon_2$. Following prior work in \cite{Dvorkin2023}, this step requires the cost $\overline{c}$ of the most expensive generator. Finally, Step 3 post-processes the initial synthetic load by solving the bilevel optimization problem \eqref{eq:pp_cro} using the conservative RO approximation of the attack. Since Step 3 does not optimize over real loads, it does not introduce any privacy loss. The complete formulation of \eqref{eq:pp_cro} can be seen in Appendix B.


The resilience of the source grid to load redistribution attacks is controlled by the parameter $\beta$ and admissible set $\Delta$. Naturally, a larger $\beta$ and a larger set $\Delta$ lead to greater resilience, but at the expense of the fidelity of the synthesized data. Our experiments in Sec. \ref{sec:experiments} will justify for the choices of these parameters. The privacy guarantee for $\alpha$-adjacent load vectors is established by the following result.

\begin{theorem}[DP of CRO]
    Setting $\varepsilon_1=\varepsilon_2=\varepsilon/2$ renders Alg. \ref{alg:CRO} $\varepsilon-$DP for $\alpha-$adjacent load vectors. \theoremsymbol \label{th:DP CRO}
\end{theorem}

\subsection{Exponential Mechanism to Ease Computational Burden}\label{subsec:exponential_mech_extension}

While the RO approximation \eqref{prob:RO} leads to a more tractable bilevel optimization, it is still computationally expensive in large systems due to the massive amount of variables and complementarity constraints, as later substantiated by Fig. \ref{fig:complexity}. We propose to alleviate the computational burden by selecting only a subset $\mathcal{K}=\{k_i\}_{i=1}^{\tau}$ of $\tau$ constraints for RO reformulation that affect the OPF cost the most. The remaining constraints $\mathcal{K}':=\{1,\dots,K\}\backslash \mathcal{K}$ are enforced deterministically.  Setting $\tau=K$ leads to the full RO formulation, while $\tau<K$ leads to a reduced problem:
\begin{subequations}\label{prob:RO_tau}
\begin{align}
    C_{\text{att},\tau}^{\text{RO}}(\mb{d})=&\;\minimize{\mb{x}}\;\; \mb{c}^{\top}\mb{x}\\
    \st\;\;\;& \underset{\mbg{\delta\in\Delta}}{\max}\left[\mb{a}_{k}^{\top}\mb{x} + \mb{b}_{k}^{\top}(\mb{d}+\bm{\delta}) + e_{k}\right] \leqslant \mb{0},\\
    & \mb{a}_{k'}^{\top}\mb{x} + \mb{b}_{k'}^{\top}\mb{d} + e_{k'}\leqslant \mb{0},\\
    &\quad\quad\quad\quad\quad\quad\quad\!\;\forall k\in\mathcal{K}, \; \forall k'\in\mathcal{K'}\nonumber. 
\end{align}
\end{subequations}

While directly replacing $C_{\text{att}}^{\text{RO}}(\tilde{\mb{d}})$ with $C_{\text{att},\tau}^{\text{RO}}(\tilde{\mb{d}})$ in Alg. \ref{alg:CRO} alleviates the computational burden, this also degrades the privacy guarantee of Theorem \ref{th:DP CRO}: since the worst-case constraint set $\mathcal{K}$ is specific to a particular load vector $\mb{d}$, the post-processing on $\mathcal{K}$ would leak information we intend to obfuscate. As a remedy, we leverage the report-noisy-max algorithm, a discrete version of the exponential mechanism of DP \cite{Dwork2014}, to privately identify the worst-case constraints without leaking information about the actual load. The resulting algorithm, termed CRO-Exp, is given in Alg. \ref{alg:CRO-Exp}.

The first two steps of Alg. \ref{alg:CRO-Exp} follow those in Alg. \ref{alg:CRO}. At Step 3, the algorithm applies the exponential mechanism $\tau$ times to construct set $\mathcal{K}$. In each iteration $t$, the mechanism identifies the constraint $k_{t}$ that---when reformulated in a robust fashion---leads to the greatest increase of OPF cost. After $\tau$ iterations, set $\mathcal{K}$ contains $\tau$ worst-case constraints. 
Finally, Step 4 solves the post-processing optimization with only $\tau$ constraints reformulated in RO way. 

\begin{theorem}[DP of CRO-Exp]\label{th:CRO_EM}
 Setting $\varepsilon_1=\varepsilon_2=\varepsilon/3$ and $\varepsilon_3=\varepsilon/(3\tau)$ renders Alg. \ref{alg:CRO-Exp} $\varepsilon$-DP for $\alpha$-adjacent loads. \theoremsymbol\label{th:DP FCRO}
\end{theorem}

\begin{algorithm}[t]
  \small
  \KwIn{$\mb{d}$, $(\alpha,\varepsilon_1,\varepsilon_2,\varepsilon_3)$, $(\beta,\gamma,\Delta,\tau)$,  $\mathcal{K}=\{\emptyset\}$
  }

  {\footnotesize \textbf{1}} Initial load obfuscation: $\tilde{\mb{d}}^{0}=\mb{d} + \text{Lap}\left(\!\frac{\alpha}{\varepsilon_1}\!\right)^{n}$

  {\footnotesize \textbf{2}} DP estimation of OPF costs: $\tilde{C}_\text{opf}=C_\text{opf}(\mb{d})+\text{Lap}\left(\frac{\alpha\overline{c}}{\varepsilon_2}\right)$

{\footnotesize \textbf{3}} DP estimation of the set $\mathcal{K}$ of the worst-case constraints

    \For{$t = 1,\dots,\tau$}{
    \For{$k = 1,\dots,K$}{
    $ C_k = C_{\text{att},t}^{\text{RO}}(\mb{d})+\text{Lap}\left(\frac{\alpha\overline{c}}{\varepsilon_3}\right)$
    }
    $k_t \leftarrow \text{argmax}_{k}\; C_k$\\[2pt]
    $\mathcal{K} \leftarrow \mathcal{K} \cup \{k_t\}$\\
    }

  {\footnotesize \textbf{4}} Post-processing optimization of the synthetic load vector
  \begin{align}\label{eq:pp_cro_exp_mech}
       \tilde{\mb{d}} \in \underset{\tilde{\mb{d}}}{\text{argmin}}\;& \lVert C_\text{opf}(\tilde{\mb{d}}) - \tilde{C}_\text{opf}  \rVert_1 \nonumber \\[-5pt]
       &\hspace{0.5em} + \beta\lVert C_{\text{att},\tau}^{\text{RO}}(\tilde{\mb{d}})-\tilde{C}_\text{opf} \rVert_1 + \gamma \lVert \tilde{\mb{d}}-\tilde{\mb{d}}^0 \rVert_1
    \end{align}
    
    \KwOut{Synthetic load vector $\tilde{\mb{d}}$}
  \caption{Privacy-preserving CRO-Exp}\label{alg:CRO-Exp}
\end{algorithm}

\section{Experiment Results}\label{sec:experiments}

We run experiments using standard power grid testbeds. The set of admissible attacks includes the limits on attack magnitude as percentage $\eta$ of nominal loads. The privacy loss $\varepsilon=1$, and we vary adjacency $\alpha$ throughout the experiments. The code and data to replicate our results are available at 
\begin{center}\small 
    \texttt{\href{https://github.com/Wu-ShengY/CRO_SynDataset}{https://github.com/Wu-ShengY/CRO\_SynDataset}}.
\end{center}

\subsection{Substantiating Attacks Calibrated on DP Data}
Table \ref{tab:BO att} collects the damage of load redistribution attacks. The synthetic loads are generated using the standard post-processing \eqref{eq:blpp} with no cyber resilience guarantee. The results reveal that the load redistribution attacks are as effective on synthetic loads as on the original loads, motivating the cyber resilient obfuscation by means of Alg. \ref{alg:CRO} and \ref{alg:CRO-Exp}. 


\begin{table}
\centering
\caption{Average outcomes of load redistribution attacks [\$ 1,000]}
\label{tab:BO att}
\begin{tabular}{lccccc}
\toprule
\multicolumn{1}{l}{\multirow{2}{*}{Testbed}} & \multicolumn{1}{c}{\multirow{2}{*}{Load}} & \multirow{2}{*}{$C_\text{opf}$} & \multicolumn{3}{c}{$C_\text{att}^{\text{BO}}$ (for varying $\eta$)} \\
\cmidrule(lr){4-6} 
& &  & $\pm5\%$  &  $\pm10\%$  & $\pm15\%$ \\
\midrule
\multicolumn{1}{l}{\multirow{2}{*}{5\_pjm}} 
& actual, $\mb{d}$          & 88.2 & 92.5 & 100.0& 108.1\\
& synth., $\tilde{\mb{d}}$  & 87.4 & 92.4 & 100.0& 108.1\\
\cmidrule(lr){1-6} 
\multicolumn{1}{l}{\multirow{2}{*}{14\_ieee}} 
& actual, $\mb{d}$          &4.80 &  4.93& 5.06& 5.19 \\
& synth., $\tilde{\mb{d}}$  &4.78 & 4.93& 5.03& 5.17\\
\cmidrule(lr){1-6} 
\multicolumn{1}{l}{\multirow{2}{*}{24\_ieee}} 
& actual, $\mb{d}$          & 227.2&  255.0& 283.0& 311.1\\
& synth., $\tilde{\mb{d}}$  & 212.5& 242.3& 259.1& 274.9\\
\cmidrule(lr){1-6} 
\multicolumn{1}{l}{\multirow{2}{*}{118\_ieee}} 
& actual, $\mb{d}$          & 237.0& 252.4&  256.4& 259.8 \\
& synth., $\tilde{\mb{d}}$  & 225.1& 229.1 & 238.8 & 241.0\\
\bottomrule
\end{tabular}
\end{table}

\subsection{Insights from the Small PJM 5-Bus Testbed}
We test the CRO Alg. \ref{alg:CRO} in mitigating the attack damage. We generate $1,000$ synthetic loads using the standard post-processing (PP) in \eqref{eq:blpp} and $1,000$ synthetic loads  from the CRO assuming $\eta=5\%$. The histograms of the normal and post-attack OPF costs are shown in Fig. \ref{fig:CRO_alpha}. Their range becomes wider as load adjacency (and hence the noise) increases. For the standard post-processing (PP) (top row), we observe a notable shift of the post-attack histogram to the right relative to the cost of normal operations, confirming the results from Tab. \ref{tab:BO att}. The attacks calibrated on the outcomes of the CRO algorithm result in no extra OPF cost, as the histograms of the normal and post-attack cost overlap (bottom row). Thus, when attacks are calibrated on CRO results, the adversary sees no gain from launching an attack.

Table \ref{tab:hyper} shows the impact of the trade-off parameter $\beta$ on the CRO algorithm. The load redistribution attack demonstrate notable damage when disregarding attacks in the CRO algorithm $(\beta=0)$. On the other hand, as long as $\beta$ exceeds the regularization weight $\gamma$, the source grid remains immune to attacks. This trade-off is ``flat'' as we model the linear OPF costs; we expect it to be smoother for quadratic OPF costs, which is a subject of future investigation.


\begin{figure}
    \centering
    \includegraphics[width=1.0\linewidth,trim= 0 25 0 30,clip]{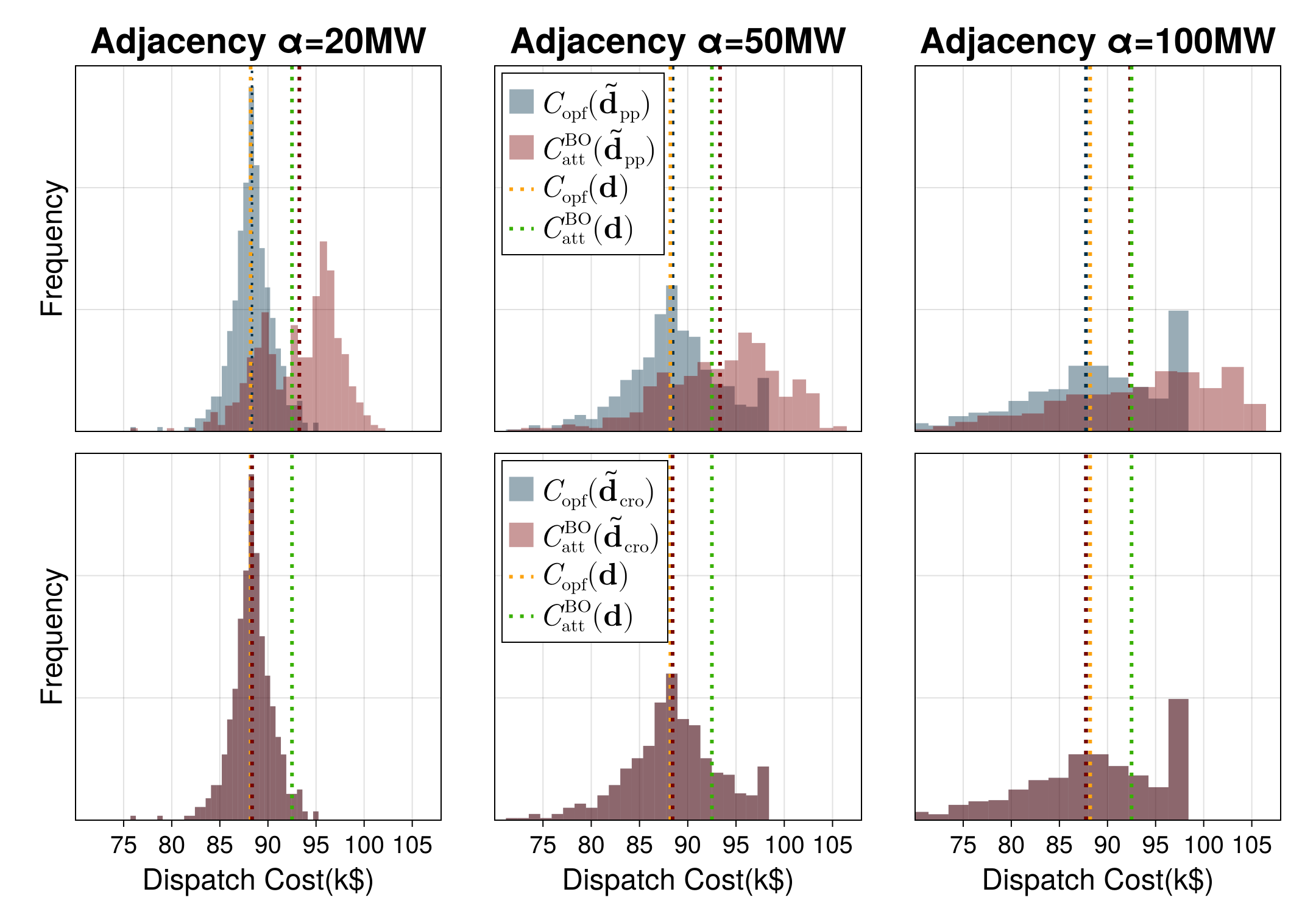}
    \caption{Histograms of normal and post-attack OPF costs in the PJM 5-bus systems. Blue and red dotted lines represent the average OPF costs on synthetic load parameters in normal and post-attack scenarios, respectively. Top row: histograms resulting from the standard post-processing based on \eqref{eq:blpp}. Bottom row: histograms resulting from the CRO algorithm.}
    \label{fig:CRO_alpha}
\end{figure}

\begin{table}[]
\centering
\caption{OPF costs induced on synthetic load vectors $\tilde{\mb{d}}_\text{cro}$ for varying trade-off parameters $\beta$ and load adjacency $\alpha$. Attack magnitude $\eta=5\%$.}
\label{tab:hyper}
\begin{tabular}{ccccccc}
\toprule
\multicolumn{1}{c}{\multirow{2}{*}{\begin{tabular}[c]{@{}c@{}}Trade-off\\ Parameters\end{tabular}}} & \multicolumn{2}{c}{$\alpha=20$ MW}  & \multicolumn{2}{c}{$\alpha=100$ MW} & \multicolumn{2}{c}{$\alpha=200$ MW} \\
\cmidrule(lr){2-3}\cmidrule(lr){4-5}\cmidrule(lr){6-7}
\multicolumn{1}{c}{}   &  $C_\text{opf}$ & $C_\text{att}^\text{BO}$  &  $C_\text{opf}$  &  $C_\text{att}^\text{BO}$   & $C_\text{opf}$    &  $C_\text{att}^\text{BO}$ \\
\midrule
  $\beta\in [0,\gamma)$ &  88.2  &  92.9   &  87.3  &  91.5 & 84.5 & 88.2 \\
  $\beta\in [\gamma,\infty)$ & 88.2 &  88.2  & 87.3 &  87.3  &  84.5  &  84.5\\
    \bottomrule    
\end{tabular}
\end{table}

\subsection{Large-Scale Applications with CRO-Exp}

The post-processing optimization \eqref{eq:pp_cro} in CRO is difficult to scale to large systems. As shown in Fig. \ref{fig:complexity}, the number of variables and complementarity constraints grow with the size of the testbed. The CRO-Exp Alg. \ref{alg:CRO-Exp} reduces the problem \revised{by at least one order of magnitude to a similar level as the standard post-processing}, since it only considers a subset of $\tau$ worst-case constraints in the attack. Fig. \ref{fig:CRO_EM_tau} shows the damage of attacks calibrated on synthetic loads released by CRO-Exp for three large testbeds. The increase of $\tau$ reduces the attack damage. Notably, $\tau=5$ suffices to minimize the attack damage, showing no improvement of cyber resilience beyond this threshold. This is due to the fact that only the attacks on a limited number of constraints can greatly increase the OPF cost. Moreover, the selection of the worst-case constraints in Step 3 of Alg. \ref{alg:CRO-Exp} becomes less informative with more noise, which only increases in $\tau$, as per Theorem \ref{th:CRO_EM}.

\begin{figure}
    \centering
    \includegraphics[width=1.0\linewidth,trim=20 20 10 40,clip]{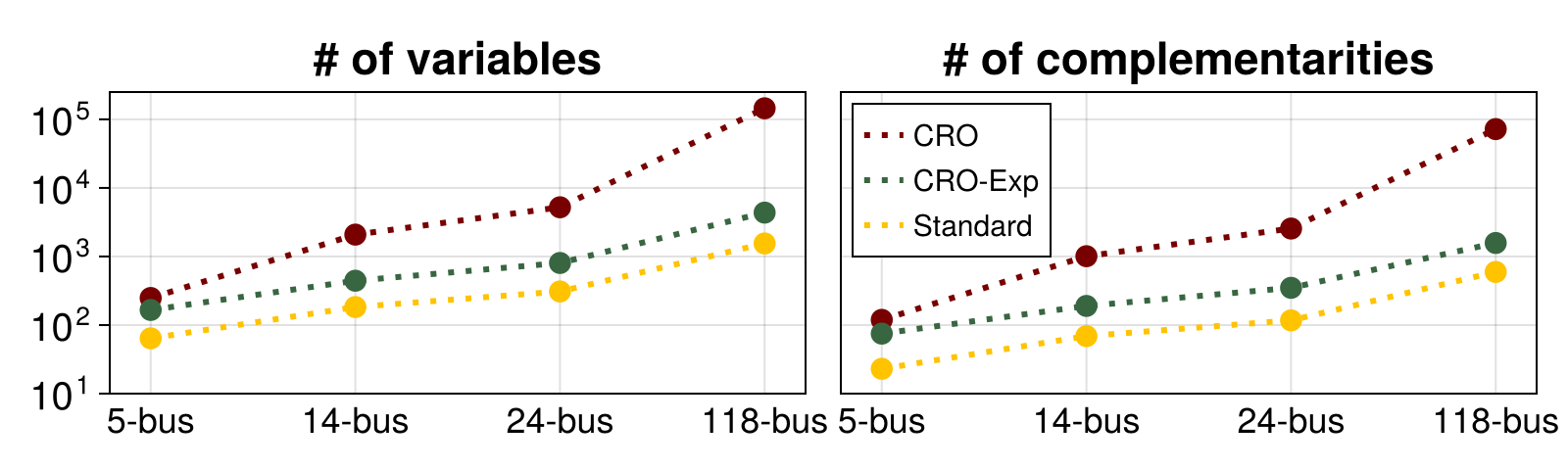}
    \caption{Num. of variables and complementarity constraints \revised{in CRO, CRO-Exp ($\tau$=5) and standard post-processing \eqref{eq:blpp}} across four testbeds (log-scale).}
    \label{fig:complexity}
\end{figure}

\begin{figure}[!tb]
    \centering
    \includegraphics[width=1.0\linewidth,trim= 0 2 0 2,clip]{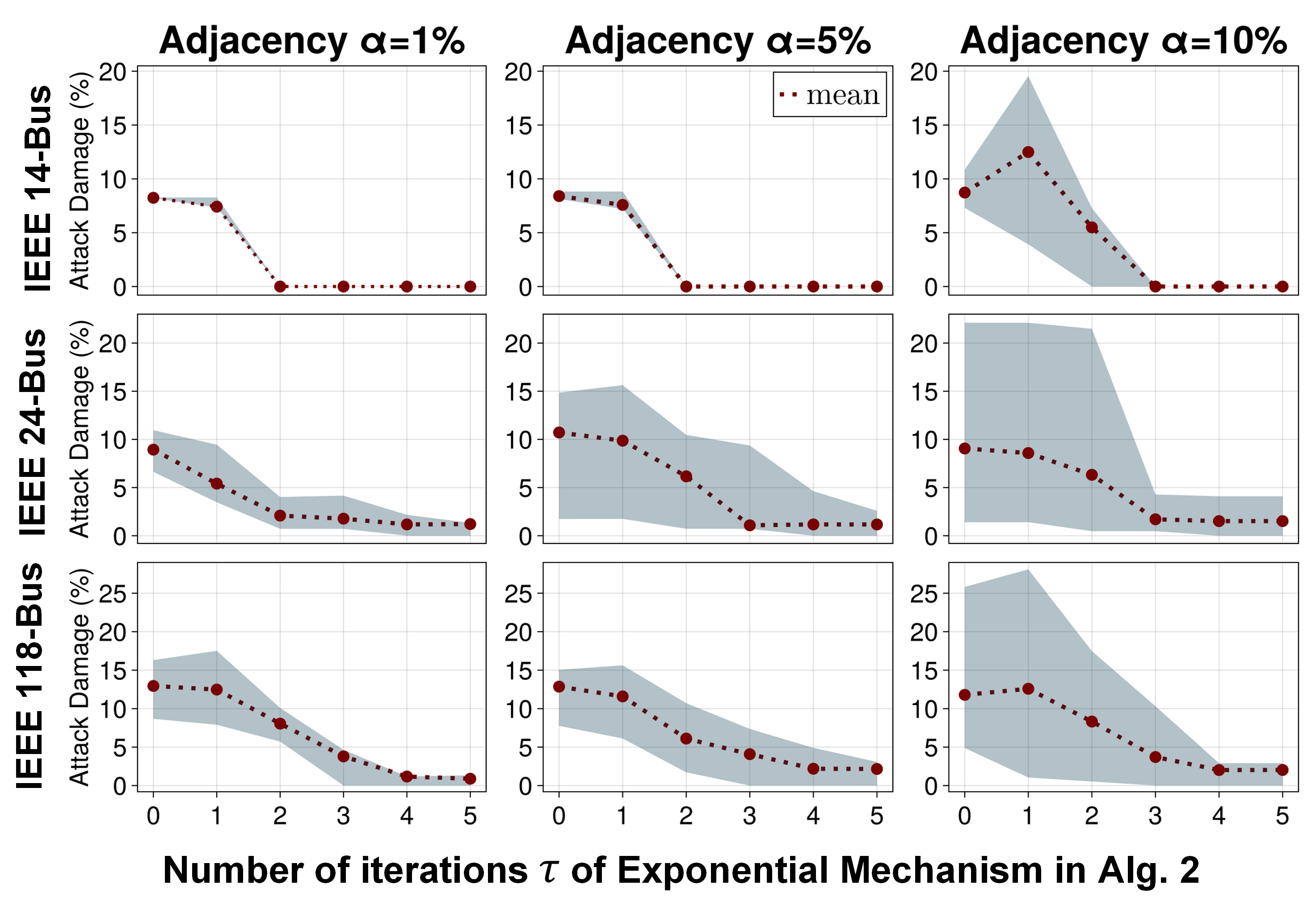}
    \caption{Outcomes of the BO load redistribution attack calibrated on synthetic CRO-Exp loads for varying number of the worst-case constraints $\tau$. \revised{The damage in percentage is computed as $(C_\text{att}^\text{BO}(\tilde{\mb{d}})-C_\text{opf}(\tilde{\mb{d}}))/C_\text{opf}(\tilde{\mb{d}})\times 100$}. $\tau=0$ means the synthetic dataset generated by the standard post-processing in \eqref{eq:blpp}. Adjacency $\alpha$ are determined in percentages of the average load in the testbed. Attack magnitudes are $\eta=15\%$ in IEEE 14-bus system and $\eta=5\%$ in IEEE 24-bus and 118-bus systems. Red lines represent the mean value, the blue area represents the 80\% confidence interval.}
    \label{fig:CRO_EM_tau}
\end{figure}

\section{Conclusion}\label{sec:conclusions}
We developed algorithms for synthesizing credible grid parameters from real-world systems for OPF analysis. Similar to existing DP algorithms, they obfuscate loads by injecting Laplacian noise and using post-processing; however, they differ in a post-processing stage which optimizes for the trade-off between modeling fidelity (OPF cost consistency) and the resilience of source grids to cyber attacks. Our results reveal that these trade-offs are ``flat'', meaning resilience can be achieved with little to no impact on the fidelity of the synthetic data. We also found that the post-processing formulation can be reduced with no loss of resilience using the exponential mechanism to select only important constraints for the attack. Inspired by these observations, future work aims to further investigate these trade-offs in the OPF setting with nonlinear (quadratic) costs and a broader class of attack models amenable to optimization-based representation.


\appendix
\subsection{Proof of Proposition 3} \label{app:att proof}

\revised{Consider two perturbed OPF problems formulated on the same vector $\mb{d}$, one resulting from the BO attack \eqref{eq:BiAttack}
\begin{subequations}
\begin{align}
    C_\text{opf}^{\text{BO}}(\mb{d})&=\minimize{\mb{x}}\quad \mb{c}^{\top}\mb{x} \label{eq:uni attack} \\[-2pt]
    \st\quad&
     \mb{a}_{k}^{\top}\mb{x} \leqslant - \mb{b}_{k}^{\top}(\mb{d}+\bm{\gamma}^{\star}) - e_{k},\;\forall k  \label{eq:opf_pert_bo}
\end{align}
\end{subequations}
and one from the RO approximation of the attack in \eqref{prob:RO}
\begin{subequations}
    \begin{align}
        C_\text{opf}^{\text{RO}}(\mb{d})&=\minimize{\mb{x}}\quad \mb{c}^{\top}\mb{x} \label{eq:pc attack}\\[-2pt]
        \st\quad&
         \mb{a}_{k}^{\top}\mb{x} \leqslant -\mb{b}_{k}^{\top}(\mb{d}+\bm{\delta}_k^{\star}) - e_{k},\;\forall k  \label{eq:opf_pert_ro}
    \end{align}
\end{subequations}
with perturbations $\bm{\gamma}^{\star},\bm{\delta}_1^{\star},\dots,\bm{\delta}_K^{\star}\in\Delta.$

To show that the optimal value of \eqref{eq:uni attack} is upper-bounded by the optimal value of \eqref{eq:pc attack}, we need to establish that the feasible set \eqref{eq:opf_pert_ro} is a subset of \eqref{eq:opf_pert_bo}. This is per the global inequality in perturbation analysis of convex programs \cite[\S 5.6, Eq. (5.57)]{Boyd2004}. Inspecting \eqref{eq:opf_pert_bo} and \eqref{eq:opf_pert_ro}, observe that this is the case when 
\begin{align}
    \mb{b}_k^{\top}\bm{\delta}_k^{\star} \geqslant \mb{b}_k^{\top}\bm{\gamma}^{\star},\quad\forall k=1,\dots,K.\label{eq:ineq_opt}
\end{align} 
The attack vectors $\bm{\delta}_1^{\star},\dots,\bm{\delta}_K^{\star}$ come from the RO, so the left-hand side of \eqref{eq:ineq_opt} is given by the following optimization: 
\begin{align}
\mb{b}_k^{\top}\bm{\delta}_k^{\star}=\max_{\bm{\delta}_{k}\in\Delta}\; \mb{b}_k^{\top}\bm{\delta}_k,  \quad \forall k=1,\dots, K. \label{eq:opt_delta}
\end{align}
At the same time, the right-hand side of \eqref{eq:ineq_opt} can be represented by the following optimization problem:
\begin{subequations}\label{eq:opt_gamma}
    \begin{align}
    \mb{b}_k^{\top}\bm{\gamma}^{\star} =\max_{\bm{\gamma}_{k}\in\Delta}\;&\mb{b}_k^{\top}\bm{\gamma}_{k} \label{eq:obj_gamma}\\[-10pt]
    &\quad\quad\quad\quad\quad\forall k=1,\dots, K\nonumber\\[-8pt]
\text{s.t.}\;&\bm{\gamma}_{k}=\bm{\gamma}^{\star}\label{eq:st_gamma}
    \end{align}
\end{subequations}
Although trivial, this optimization problem allows us to clearly relate both sides of inequality \eqref{eq:ineq_opt} by relating problems \eqref{eq:opt_delta} and \eqref{eq:opt_gamma}. They are similar except for the additional consensus constraint \eqref{eq:st_gamma}. Since $\bm{\gamma}^{\star}\in\Delta$ by design, the feasible set of $\mbg{\gamma}_{k}$ is the subset of that for $\mbg{\delta}_k$. Hence, we can conclude that the optimal value of \eqref{eq:opt_delta} is greater or equal than that of \eqref{eq:obj_gamma}. Therefore, inequality \eqref{eq:ineq_opt} holds and  \eqref{eq:opf_pert_ro} is indeed a subset of \eqref{eq:opf_pert_bo}, completing the proof.}

\subsection{Complete Formulation of the CRO Post Processing}
The complete formulation of \eqref{eq:pp_cro} with the Karush-Kuhn-Tucker conditions (KKTs) of embedded problems is
\begin{subequations}
\begin{align*}
    \minimize{}\;& \lVert \tilde{C}_{\text{opf}}-\mb{c}^{\top}\mb{x}_{1} \rVert_1 + \beta\lVert \tilde{C}_{\text{opf}}-\mb{c}^{\top}\mb{x}_{2} \rVert_1 + \gamma\lVert \tilde{\mb{d}}-\tilde{\mb{d}}^0 \rVert_1 \\ 
    \st \;\;&\nonumber\\[-10pt]
    \raisebox{-6.5em}{\rotatebox{90}{\tcbox[size=fbox]{KKTs of RO approxiamtion \eqref{prob:RO_ref}}}} &\left\{
    \begin{array}{ll}
    \overline{\bm{\mu}}_k^{\top}\overline{\bm{\delta}}-\underline{\bm{\mu}}_k^{\top}\underline{\bm{\delta}} \leqslant -\mb{a}_{k}^{\top}\mb{x}_1-\mb{b}_k^\top \tilde{\mb{d}}-e_k \\
         \mb{b}_k-\overline{\bm{\mu}}_k+\underline{\bm{\mu}}_k-\bm{1}\lambda_k=\bm{0} \\
         \underline{\bm{\mu}}_k,\overline{\bm{\mu}}_k \geqslant \bm{0}\\
         \mb{c}+\sum_{k}\theta_{k} \mb{a}_{k}=\mb{0}\\
         \theta_{k}\overline{\bm{\delta}}-\bm{\zeta}_{k}-\overline{\bm{\pi}}_k=\bm{0}\\
         \theta_{k}\underline{\bm{\delta}}-\bm{\zeta}_{k}+\underline{\bm{\pi}}_k=\bm{0}\\
         \bm{1}^{\top}\bm{\zeta}_{k}=0 \\
         0 \leqslant \theta_k \;\perp (-\overline{\bm{\mu}}_k^{\top}\overline{\bm{\delta}}+ \underline{\bm{\mu}}_k^{\top}\underline{\bm{\delta}}- \mb{a}_{k}^{\top}\mb{x}_1 \nonumber\\
        \quad\quad\quad\quad\quad\! - \mb{b}_k^\top \tilde{\mb{d}}-e_k) \geqslant 0\\
         \bm{0} \leqslant\underline{\pi}_k \;\perp \underline{\bm{\mu}}_k \geqslant \bm{0} \\
         \bm{0} \leqslant\overline{\pi}_k \;\perp \overline{\bm{\mu}}_k \geqslant \bm{0}\\
         \theta_k,\overline{\bm{\pi}}_k,\underline{\bm{\pi}}_k,\bm{\zeta} \geqslant \bm{0}
        \end{array}
        \right\}\\[-10pt]
    \raisebox{-3.5em}{\rotatebox{90}{\tcbox[size=fbox]{KKTs of OPF \eqref{eq:DCOPFc}}}}&\left\{
    \begin{array}{ll}
         \mb{a}_{k}^{\top}\mb{x}_2+\mb{b}_k^\top \tilde{\mb{d}}+e_k \leqslant 0 \\
         \mb{c}+\sum_{k}\nu_{k} \mb{a}_{k}=0 \\
         0 \leqslant \nu_k \;\perp (- \mb{a}_{k}^{\top}\mb{x}_2 - \mb{b}_k^\top \tilde{\mb{d}}-e_k) \geqslant 0\\
         \nu_k \geqslant 0
    \end{array}
        \right\}
\end{align*} 
\end{subequations}
$\forall k=1,\dots,K$, where $\mb{c}^\top\mb{x}_1$ and $\mb{c}^\top\mb{x}_2$ represents the OPF cost in post-attack and normal conditions, respectively. Here, the $\perp$ denotes complementarity conditions. 
\subsection{Proof of Theorem \ref{th:DP CRO}}
CRO uses the real data in the following computations:
\begin{enumerate}
    \item Step 1 adds Laplacian noise with magnitude $\alpha/\varepsilon_2$ to an identity query,  whose sensitivity is $\alpha$. By the sequential composition rule \cite{Dwork2014}, this computation is $\varepsilon_1-$DP.
    \item Step 2 adds Laplacian noise with parameter $(\alpha \overline{c})/\varepsilon_2$. Since the sensitivity of OPF cost is $\alpha \overline{c}$ as shown in Section II.B, this computation is $\varepsilon_2-$DP.
\end{enumerate}
Since the post-processing optimization in Step 3 only uses obfuscated data, it will not induce any privacy loss due to post-processing immunity \cite{Dwork2014}. Per the sequential composition rule, the total privacy loss of the algorithm is $\varepsilon_1+\varepsilon_2$, which adds up to $\varepsilon$ when we take $\varepsilon_1=\varepsilon/2, \varepsilon_2=\varepsilon/2$.

\subsection{Proof of Theorem \ref{th:DP FCRO}}
The algorithm queries data in the following computations:
\begin{enumerate}
    \item Following the similar arguments from Appendix C, Step 1 is $\varepsilon_1-$DP and Step 2 is $\varepsilon_2-$DP
    \item The worst-case constraints are estimated using $\tau$ iterations of the report-noisy-max algorithm in Step 3; each iteration injects the Laplacian noise with magnitude $\alpha \overline{c}$ providing  $\varepsilon_3-$DP, and the whole report-noisy-max algorithm is $\tau\varepsilon_3-$DP.
\end{enumerate}
As the post-processing optimization in Step 4 only uses obfuscated numerical data $\tilde{\mb{d}}^0, \tilde{C}_{\text{opf}}$ and non-numerical data $\mathcal{K}$, it is immune to privacy loss. The accumulated privacy loss of Alg. \ref{alg:CRO-Exp} is $\varepsilon_1+\varepsilon_2+\tau\varepsilon_3$, which amounts to $\varepsilon$ when setting $\varepsilon_1=\varepsilon/3, \varepsilon_2=\varepsilon/3$ and $\varepsilon_3=\varepsilon/(3\tau)$.



\bibliographystyle{IEEEtran} 
\bibliography{CDC2025}

\endgroup
\end{document}